# Fritz Haber, the damned scientist

Magda Dunikowska[*] and Ludwik Turko[+]

[*]*e-mail:* magda_dunikowska@hotmail.com
[+] *Institute of Theoretical Physics, University Wrocław, pl. M. Borna 9, 50-205 Wrocław, Poland*
*e-mail:* turko@ift.uni.wroc.pl

**A portrait and a monograph**

The even row of portrait photographs of Lower Silesian Nobel Prize winners displayed on the wall of the club Salon Śląski, or Silesian Salon, one of the city's magic places, right across the street from the Baroque main building of the Wrocław University, is rather unorthodox as far as the standards of picture exhibitions go. Two of the laureates observe the cosy interior of the club having assumed postures that are somewhat unusual for respectable learned men: hanging upside down. One of the two is Philipp Lenard, the cathode ray discoverer who subsequently developed the conception of creative 'Aryan physics' as opposed to secondary and mendacious 'Jewish physics'. The other one is Fritz Haber, who invented a method for synthesizing ammonia and later pioneered the use of poison gases on World War I battlefields. In the gallery of famous people tracing their origins to Wrocław, few are as controversial, as complex, or as tragic as Fritz Haber. He was awarded the Nobel Prize in chemistry in 1919 for developing a method for the direct synthesis of ammonia from its elements: hydrogen and nitrogen. The reaction made possible industrial-scale production of artificial fertilizers to provide grain crops with necessary nitrogen. For hundreds of millions around the world, the discovery averted the spectre of famine and linked Haber's name with the concept of 'bread from air'. It would be difficult to find a better illustration of Alfred Nobel's last will, which instructed his heirs to bestow prizes on those who confer the greatest benefit on mankind.

Less than a decade after enabling the production of bread from air, Fritz Haber pioneered the use of deadly poison gases on the battlefields of World War I. He personally oversaw the first successful chlorine gas attack on the French and English lines at Ypres in April 1915. His passion and commitment led to the association of Haber's name with the notion of 'poison from air'.

The authors of *Microcosm: Portrait of a Central European City*, Norman Davies and Roger Moorhouse, dot the i's and cross the t's: 'Fritz Haber (1868–1934) ... earned the name of Germany's "Doctor Death". After studying in Berlin, he returned to Breslau to take over his father's business, but tired of merchant life and opted for an academic career. Though largely self-taught, he lectured at the Technical Highschool in Karlsruhe before being appointed Professor of Physical Chemistry ... At the outbreak of war in 1914, he placed the institute at the disposal of the government and became involved in the development of chemical weapons. Less than a year later, on 22 April 1915, Haber personally directed the German chlorine gas attack at Ypres. His wife and fellow chemist, Clara Immerwahr, committed suicide in protest at his work, but he pressed on undeterred. He was later to be involved in the development of "Zyklon B".'[1]

In the light of the above, it would seem quite fair and proper to hang the Germany's 'Doctor Death', in effigy, not only upside down but also facing the wall. Before doing that, however, and before embarking on an anti-Haberian crusade, replete with easy moralistic indignation,[2] it is worthwhile taking a closer look at this character, in whose story crystallized the key challenges and phobias of his time. To begin with, it is reasonable to put aside *Microcosm*, at least as a source of knowledge about Fritz Haber. Describing a graduate of the University of Heidelberg with a Ph.D. in chemistry from Berlin as 'self-taught' is rather precarious, at least as much as is calling the Technische Hochschule Karlsruhe a high school. It would be as appropriate and informative to call Poland's Szkoła Główna Handlowa a trade high school or the École Normale Supérieure in Paris an ordinary high school.

In actual fact, this scientist who has become a black legend, his decisions, and his fate deserve a fair and objective analysis for at least two reasons. Firstly, because of the character, talent, and achievements of this extremely complex personality who was a true hero of his time. Secondly, because such analysis provides an opportunity to gain an insight into the beginnings of the era that turned scientists and industrialists into new political players, i.e. our present time. The figure of Fritz Haber, like a lens, brought into focus all of the tough dilemmas of abandoning the romantic vision of history, still alive during his lifetime. Let us treat him then as a window into the Wrocław/Breslau and the Europe of that time with their conflicts, hopes, and achievements, and into the point where a paradigm shift took place marking one of the major civilizational turning points: the world would never be the same after Haber's inventions; much like the world would never be the same after the breakthroughs of his friend Einstein. Today's landscape with millions of shops selling fresh packaged foodstuffs, restaurants and fast-food outlets mushrooming on all continents and even the most remote islands, the landscape that is, as it were, our natural environment, has come into being as a consequence of none other than Haber's work.

Fritz Haber, a true Breslauer by birth, grew up in a city that was a European microcosm. *Microcosm,* the title chosen by the authors for the above-cited monograph of the city, aptly captures the essence of the place, including especially the fervour of late 19th century Breslauers. The city, a mixture of ethnicities, cultures, and religions, was torn between the poles of elegant urban culture and faith in the power of science. Its outwardly manifest growth proceeded in a climate of immediate industrial-scale application of chemical patents, which was made possible by the collaboration of university laboratories, but in a way evoked echoes of alchemy. The murky yearning for power promised affluence, unmindful of the risk of unleashing forces that could spiral out of control and push the world into the turmoil of destruction. The city, rapidly growing ever prettier, seemed to be inhabited by a *genius loci*, a kind of guardian spirit of the place, protecting its residents. The *genius loci* may have been present at the bed of a certain Breslau woman in labour who was giving birth to Fritz Haber, the father of weapons of mass destruction but also of a technology used to avert mass famine.

---

[1] Norman Davies and Roger Moorhouse, *Microcosm: Portrait of a Central European City* (London, Jonathan Cape, 2002)
[2] See, e.g., Mariusz Urbanek, 'Żona Doktora Śmierć' [Doctor Death's wife], *Gazeta Wyborcza (Wysokie Obcasy),* 8 January 2009; Gerit von Leitner: *Der Fall Clara Immerwahr: Leben für eine humane Wissenschaft* (C.H. Beck Verlag 1996)

Born into a Jewish family tracing its roots to Galicia, an area spanning today's south-eastern Poland and western Ukraine, Fritz Haber was an intellectual with a posture and personality of a Prussian *Junker*. His ammonia synthesis not only made it possible to mass-produce artificial fertilizers but also enabled industrial production of compounds needed to mass-produce explosives. Haber was a fierce German patriot at a time when state nationalism was a virtue and a commendable attitude. After the outbreak of World War I, he was convinced that the shock caused by chemical weapons would force the Entente to quickly capitulate, thus saving lives. That is where he was wrong: for over three years millions of soldiers would continue to die in the muddy trenches of the Great War. Chemical weapons, used by all the belligerent countries, did not bring about any breakthrough, and 'traditional' weapons were much more efficient in killing people than the chemicals. The latter would not prove their superior efficiency until their application in German death camps during World War II.

After Hitler's rise to power, state nationalism was supplanted by ethnic nationalism and the German Haber became the Jew Haber. A year later, having left Germany, he died in Basel. At a semi-conspiratorial memorial service held at the Kaiser Wilhelm Wilhelm Society, whose Institute for Physical Chemistry and Electrochemistry Haber had directed since 1911, another German Nobel Prize winner, Max Planck, stressed that without Haber's work on ammonia synthesis Germany would have lost World War I after just a few months both for economic reasons, lack of food, and for military reasons, lack of ammunition. It so happens that the reaction providing bread from air also makes it possible to produce explosives. Planck's speech was delivered to a tightly packed audience, mostly composed of women, professors' wives. They were representing their husbands, who preferred to stay at home choosing 'the lesser evil' and 'preservation of values'.

**Understanding Haber**

There are multiple roads to understanding the extraordinary personality of Fritz Haber. Travelling those roads are numerous contemporary historians, biographers, film makers, and artists.[3] Haber's name appears in theatre plays, novels and biographies[4]. Just how it continues to intrigue and inspire to this day is evidenced by the *Fritz Haber* series started a few years ago. To its creator, David Vandermeulen, a talented Belgian graphic artist, this complex character brings into focus the complexities of the early industrial era: the dynamics of brilliant technological advances fuelled by the ambitions of newly formed social classes. He even developed a special literary genre for his protagonist: an interesting combination of comic book, drama, and historical documentary. The resulting opus is a book/album/portal in sepia, resembling a silent motion picture, where dialogues alternate with situational descriptions.[5] The author has taken particular care to ensure that his work has historical value and put it in a different perspective. In addition to classic quotes from acclaimed authors, chapters open with longer quotations from period documents: speeches by presidents, prime ministers, generals, statements by leading journalists and writers, i.e. various opinion leaders, published in European newspapers and magazines of that time. It is the beginning of a literary/artistic/research project planned for more than ten years: this is how much time the author expects is needed to sort out the tragic destiny of this Faust of the turn of the twentieth century.

---

[3] See, e.g., Margit Szöllösi-Janze, *Fritz Haber, 1868–1934* (München, C.H. Beck, 1998); Dietrich Stoltzenberg, *Fritz Haber: Chemiker, Nobelpreisträger, Deutscher, Jude* (Weinheim, Wiley-VCH, 1998), English translation: *Fritz Haber: Chemist, Nobel Laureate, German, Jew* (Philadelphia, Chemical Heritage Foundation, 2006); Daniel Charles, *Master Mind: The Rise and Fall of Fritz Haber, the Nobel Laureate who Launched the Age of Chemical Warfare* (New York, Ecco, 2005); F. Stern, `Together and Apart: Fritz Haber and Albert Einstein`, *Einstein's German World: Essays in European History* (Princeton University Press, Princeton, 1999), *Dreams and Delusions: The Drama of German History* (Columbia University Press, New York, 1987); M. Goran, 'The Present-Day Significance of Fritz Haber', *American Scientist*, 35:3 (1947), pp. 400–403; *The Story of Fritz Haber* (University of Oklahoma Press, 1967); Daniel Ragussis (dir.), *Haber,* short film, 2008;

[4] For a critical review see, e.g., Roald Hoffmann and Pierre Laszlo, 'Coping with Fritz Haber's Somber Literary Shadow', *Angew. Chem. Int. Ed. Engl.* **2001**, 40, pp. 4599 – 4604; Bretislav Friedrich, *Angew. Chem. Int. Ed. Engl.* **2005**, 44, pp. 3957-3961 and **2006,** 45, pp. 4053-4055

In the face of so many manifestations of ongoing interest, one cannot do justice to the Fritz Haber figure other than by recognizing three layers of its structure: internal, or the psychology of his personality; external, or his social status; and temporal, or his development and evolution driven by family history and social pressures.

A review of the literature on Fritz Haber offers an insight into the impressively complex landscape of his time. On the one hand, the scholarly approach makes possible a solid reconstruction of the industrial era on the eve of World War I, when it had already become clear that it would be very difficult for an inventor to stop at being a benefactor of mankind. No society can restrict, or could have restricted, a new, promising technology to serve peaceful development only. On the other hand, the humanities will not allow the ethical questions to be left out, bringing up the issue of scientists' responsibility. Likewise, it is impossible to pass over the role of pressures from various interests which, like a powerful tributary, bolster social dynamics and their further turbulent development. Thus, it is difficult to ascribe all further uses of an invention to the will of the inventor. Consequently, the story of Fritz Haber does not permit an account composed of simple constatations, forcing the analyst to suspend judgments and concentrate on the question marks.

**The virtues of patriotism**

If we take a really close look at the time of Haber's youth in order to investigate German cities' intellectual climate *in statu nascendi*, to trace the ambitions of the elites of Bismarck's state, whom will we meet? The world in which Fritz Haber grew up and was educated was already very complex. Prussia laid emphasis on solid and rigorous education, with discipline, patriotism, and respect for the army instilled at home and at school. As the education was comprehensive and of high quality in every field with the purpose of ensuring cohesion of the state, the nationalism emerging in those conditions did not appear in the least pathological, especially as regional and religious differences were still visible and the bloody civil wars were still present in living memory. It was that memory that Wilhelm II and his chancellor tried to console by uniting the German peoples under the new motto of 'Deo – Litteris – Patriae'. Scholars and scientists could hope for a high rank in the social hierarchy. This is how the *Kaiser* congratulated Wilhelm Röntgen in 1896 on the discovery of X-rays: 'I praise God for granting our German fatherland this new triumph of science.' Thus, Haber's immediate environment was marked by fresh dynamics of a sensibly developing state. The emblematic trio of God, Science, and Fatherland seemed in a natural way to assure the right course of civilizational evolution. An atmosphere of confidence in the virtues practised set in, especially as the appearance of cities clearly manifested not only wealth but also beauty and harmony.

---

[5] http://www.editions-delcourt.fr/fritzhaber/

Commerce and industry coupled with love of the army did not eliminate love of art, which was revealed in architecture, town planning, sculpture, painting, and handicrafts. If one makes the effort to reconstruct the streets that Haber walked, the laboratories, lecture halls, and salons he spent time in, and the furnishings and thousands of objects he used, it is not difficult to realize that all of that outside world was in fact part of his world. It certainly gave him a lot of satisfaction: he felt at home, at the right place, ready to work incessantly in order to maintain that state of affairs.

By the end of the nineteenth century, industrial development had attained unprecedented intensity, due to constant rivalry between the states of Europe. It soon turned out that the mutual stimulation came at a price: nationalism was rising, and with it, in view of conflicting interests and ambitions of the European powers, grew the threat of war. Under such conditions, for many, patriotism imperceptibly mingled with nationalism, so much so that the boundary between the two was no longer discernible. From today's perspective, following the painful experience of the paroxysms of the twentieth century, it is no longer possible to easily picture a time when extreme nationalism was a virtue and a commendable attitude. The latter half of the nineteenth century was a time a fierce rivalry in Europe between the French, the Germans, and the English. The united German empire, emerging from nonexistence lasting since the Thirty Years' War, the *Kaiserreich*, which considered itself the successor of the Holy Roman Empire of the German Nation, vigorously pushed and shoved to gain elbowroom in a space already divided up by Europe's traditional colonial powers. The conflicting currents of state nationalisms clashed in all areas of life, without omitting the traditional academic virtues. It was at that juncture that Pierre Duhem, an eminent French physicist and philosopher, in his essay 'La science allemande', contrasted the *esprit de finesse* of French science with Germany's dull scientific thought that he considered a degenerate form of French science. He also used the opportunity to expose the shortcomings of English scientific thought as, while not deprived of sharpness, suffering from a shortage of logical coherence, or *bon sens*.

**Between prosperity and the spectre of famine**

The modern reader rarely has a chance to take a close look at the decades preceding the outbreak of World War I through the prism of documents originating from industrial companies, university laboratories, and research institutes.

The rapid development enjoyed by Europe in the industrial age was not free of concerns. Threats to the development of the European industrial civilization were discussed since Malthus. Even as a century earlier, he had warned that advances in European civilization, which extended life expectancy and thus resulted in a steady population growth, would come up against the problem of feeding the population. Existing production, dependent on the whims of climate and the land drained by centuries of cultivation, would be unable to meet the growing demand. Mankind would thus be left with the only choice to restore a balance: famine or war. In the late nineteenth century, the problem persisted to be the main challenge to be tackled by science: Sir William Crookes, president of the British Association for the Advancement of Science, who discovered thallium and invented the radiometer, presented this issue as a potentially imminent catastrophe. Imports of American and Russian grain, which had been helping to maintain a balance, would no longer be able to fulfil the task, since those main producers would limit supplies in the coming decades in order to feed their own populations. Restoring internal self-sufficiency was not an option either: Chilean sodium nitrate deposits and reserves of South American guano were nearly exhausted. The only solution was to develop a method for the production of fertilizers from ammonia taking nitrogen directly from its inexhaustible source, i.e. air.

Crookes' speech resounded throughout the scientific communities in Europe, and 1898 marked the beginning of a race among laboratories. German scientists had a privileged position at the time as their country had introduced an innovative solution to the burning problem of research funding: government guarantees encouraged both bankers and industrialists to invest in science. Factories willingly purchased patents and employed talented specialists, while banks provided financing. The effectiveness of the resulting academic-industrial-financial complex proved disturbing for other countries.

By the turn of the twentieth century, the high status that Bismarck's reunified state afforded to the symbiosis of science and industry had seriously undermined the traditional dominance of the colonial powers. The patent race was ever more clearly tipping in favour of Germany when its scientists discovered the structure of alizarin, the main ingredient of dyer's madder. As early as 1872, synthetic alizarin was produced by three different German chemical concerns: BASF, MLB, and Bayer. This powerful competition soon ruined the traditional cultivation of madder, the cost of the synthetic dye being a tenth of that of the natural substance. Less than fifteen years was enough to see a complete collapse of the market. The south of France, which in 1881 still provided more than a half of the global production, sold none at all just five years later. A similar fate befell the English market for indigo, the king of dyes, even though developing a synthesis method took BASF and MLB chemists twenty-two years of incessant effort and consumed millions of marks of capital expenditure before a success was achieved. By 1904, Germany was exporting 9,000 tonnes of synthetic indigo, rising to three times as much in 1913. That spelt ruin, now for entire regions of British India living off the cultivation of *Indigofera* plants, and consequently brought about the collapse of the English indigo market and the port of Marseille, which served that market.[6] As can be seen, globalization is by no means a new development of the last quarter-century.

Fritz Haber's career as an eminent scientist, which started with the development of an industrial method for ammonia synthesis (the so-called Bosch-Haber process), and the establishment of a Kaiser Wilhelm Institute for Haber in Dahlem were possible precisely because of the comprehensive development of German state institutions. The first attempts at ammonia synthesis were undertaken by Wilhelm Ostwald, an eminent chemist and future Nobel Prize winner, his method however was unsuccessful. Several years later, Haber and a young English scientist, Robert Le Rossignol, achieved the first promising synthesis by using precise physicochemical analysis combined with bold engineering. This was initially carried out at the laboratory of the Technische Hochschule Karlsruhe and then, on an industrial scale, in collaboration with BASF. In July 1909, the first millilitres of ammonia containing exclusively atmospheric nitrogen, unavailable before, flowed from their tabletop laboratory apparatus.

**German Jews, Jewish Germans**

Fritz Haber never had any doubts about his national identity. He considered himself and was German. German culture was his culture; the German state, *Kaiserreich*, was his state. More than a century of Prussian enlightened absolutism, going back to Frederick the Great, had led

---

[6] F. Delamare, B. Guineau, *Les materiaux de la couleur* (Paris, Gallimard-Découvert, 2010)

to the emergence of a modern state, one of the key players on the European scene. Germanhood was the young Haber's natural environment, something 'not to judge, but to adjust to, like day and night, like spring and summer, like everything great and eternal'.[7] One of the aspects of Haber's self-fulfilment in Germanhood was his decision, at the age of twenty-four, to be baptized in a Protestant church.

For the talented and ambitious man, brought up in a country with efficiently functioning institutions, a scientific career was a powerful attractive force, stronger than his family's religious tradition. It is worth pointing out that, since the Stein-Hardenberg reforms in the late eighteenth century, the inhabitants of Prussia had been treating their state as a tool of emancipation, universal education, and formation of a sense of citizenship. Nevertheless, the emancipation of Jews was not a completed process at that time, despite successive legislative advances, and persisting anti-Semitism in nineteenth-century Europe was far from a marginal phenomenon. Thus, when Haber's religion became a serious handicap, he decided to convert to Christianity. The decision was commented upon variously, but there is no doubt that in this case, like in many similar cases, religious motives played a much less significant role than a desire to open up and assure one's career prospects. This conclusion is evidenced by events in Haber's later life, which confirmed that the main reason for his conversion was a desire to blend into Germanhood, a need to feel 'one of us', a community bound by ties of a common land, a common past, and a common present.

While changes taking place in Prussian society had gained a momentum unseen before, the growing emancipation of various social strata proceeded for the time being without the former elite being stripped of its privileges. The electoral system based on three property-owning classes, rooted in a long European tradition of membership of various occupational corporations, gave the members an important place in the social hierarchy. The Jews, invariably involved in commerce and international finance since the ancient times, had always enjoyed direct access to the monarch, which was considered a particular privilege. When Bismarck succeeded in unifying Germany towards the end of the nineteenth century, the development of the German state in the new structures, based on industrial investments and maintenance of extensive, increasingly international markets, proceeded in parallel with the rise of a new elite: Jewish bankers and industrialists. Due to its strength, that social stratum, by its very nature cosmopolitan and cultivating somewhat different codes of social communication, was perceived by Prussian aristocratic families to be a dangerous competitor that might in time become a threat to the construction of their state, based since the time of Frederick the Great on military might and a high level of education. Thus, despite formal acts of enfranchisement, the political reality revealed new divisions, fractures, and tensions. The German officer corps with traditions rooted in old *Junker* families remained an impregnable fortress, out of reach even to assimilated Jews converted to Protestantism; the university elite was similarly hermetic, carefully scrutinizing all candidates.

Fritz was entering adulthood just as another wave of discussions about the role of Jews in the new Reich was sweeping through Germany. It was not a purely German problem. The question about the place of Jews in the states of Europe in the Age of Enlightenment was a question about the practical implementation of the concept of a state that, at least in principle, afforded equal rights to all. In England, the Jewish question was the subject of a debate in the mid-eighteenth century; in post-revolutionary France the *question sur les juifs* was debated by the National Assembly in 1790. The German Reich guaranteed constitutional equality of its Jewish citizens upon its establishment in 1871. In that way, the state that was bringing unification after centuries of fragmentation became a synonym of a 'new order' in citizenship terms for the Jewish community. However, the adoption of the constitution did not, and in fact could not, remove the *Judenfrage,* the dispute about the place of Jews in Germany, from the public space.

---

[7] Fritz Haber – speech on the 50th anniversary of the Academic Literary Society, Universität Breslau, 10 June1924, published in Fritz Haber, *Aus Leben und Beruf*: *Ansätze, Reden, Vorträge* (Berlin, Julius Springer, 1927)

What probably also had significant effect on the young Haber was the fierce public debate sparked off by a famous essay by Heinrich von Treitschke, a Berlin-based historian, philosopher, and politician, deputy to the Reichstag, entitled 'Unsere Aussichten' (Our prospects) and published in 1879.[8] It was there that the notorious sentence 'Die Juden sind unser Unglück!',[9] which was to become the motto of the Nazi rag *Der Stürmer* barely fifty years later, first appeared in print. Treitschke's essay, distributed as a self-contained brochure entitled *Ein Wort über unser Judenthum,*[10] ignited a heated debate, referred to as *Treitschkiade* at the time and now usually described as the Berlin Anti-Semitism Dispute in the literature.[11] In contrast to the widely circulating and diverse anti-Semitic literature existing before, this came from a university scholar, a recognized authority in his field. The matter thus gained an additional dimension, and its weight increased. An answer from a polemicist equal to Treitschke in stature did not come until a year later: December 1880 saw the publication of a sixteen-page brochure by Theodor Mommsen, a professor of the University of Berlin, whose title, *Auch ein Wort über unser Judenthum,*[12] alluded to Treitschke's essay.

Theodor Mommsen, an expert on the history of ancient Rome and Greece who would win the Nobel Prize for literature in 1902 and who had been a professor of Breslau University in 1854–1856, was a recognized authority and a kind of *guru* of the German liberal circles of his time. An enthusiast of building a strong Germany, he saw the state as a community of various groups, social as well as ethnic, who self-limit their separate interests in the name of the state as a supreme good and contribute their best qualities to the 'German alloy'. That was a different conception of Germany from Treitsche's idea of a state based on blood ties and a mythical German spirit. One of the elements of the self-limitation and sacrifice that Mommsen proposed was, in the case of the Jewish community, baptism and conversion to Christianity. He argued that 'remaining outside the boundaries of Christendom and at the same time belonging to the [German] nation is possible, but difficult and risky.'[13]

It is unreasonable to assume that Fritz, twelve years of age at the time, was a keen reader of Treitsche's or Mommsen's writings. Still, both essays carried enough weight to be repeatedly reissued and become permanent reference points in never-ending discussions, throughout the 1880s and beyond. When the 58-year-old Fritz Haber confided in his friend Rudolf Stern in 1926 about what had prompted him to make the decision to convert, he admitted it was the Theodor Mommsen text.[14] 'I considered myself a hundred per cent

---

[8] Heinrich von Treitschke, 'Unsere Aussichten', *Preußische Jahrbücher*, 44 (1879), pp. 559–760
[9] The Jews are our misfortune!
[10] A word on our Jewry
[11] Walter Boehlich (ed.), *Der Berliner Antisemitismusstreit* (Frankfurt am Main, Insel-Verlag, 1965/1988)
[12] Another word on our Jewry
[13] *Theodor Mommsen: Gelehrter, Politiker und Literat*, Josef Wiesehöfer (ed.) and Henning Börm (Stuttgart, Franz Steiner Verlag, 2005). p. 152: 'Außerhalb dieser Schranken zu bleiben und innerhalb der Nation zu stehen ist möglich, aber schwer und gefahrvoll.'
[14] Rudolf A. Stern, 'Fritz Haber*:* Personal Recollections', *Leo Baeck Yearbook* 8 (1963), pp. 71–102

German and under the impact of philosophy and science, of the whole rational temper of the word, no longer felt any ties to the Jewish religion,' he recalled.

One can easily picture Fritz Haber reading Mommsen's conclusion: 'Entry into a great nation comes at a price. The Hanoverians, the Hessians, and we, from Schleswig-Holstein [Mommsen's native land], are prepared to pay it. We can feel that we are sacrificing a part of ourselves. But we are offering it to our common fatherland. The Jews too will not be led by a Moses again to the promised land.'[15]

Many years later, the now fully mature Fritz Haber with full conviction and a sort of pride addressed a group of American physicians visiting Berlin as follows: 'You live in a land where personal freedom is the highest good. Your tradition honours the pioneer whose happy work changed a dangerous wilderness into an industrial state which serves its citizens ... In our past times, not personal freedom but citizen organization was the highest political good. Our tradition does not honour the power to do but loyalty to duty. Our state does not serve its citizens, but the citizens the state. Therefore our Republic is different than is yours.'[16] It was already the year 1926.

Thirteen years later, in October 1939, another Breslau Jew, Willy Cohn, a historian deprived of work, living in a city of vandalized synagogues, reduced to second-class citizenship with a passport stamped by the police, wrote in his systematically kept diary:[17] 'I have read the Führer's speech. It was fairly moderate and reasonable; it could even be a bridge to peace if others were reasonable. But it is unlikely that England will acquiesce. The speech was not particularly anti-Semitic, either. One should acknowledge the greatness of the man who has given the world a new face.' That ethos of Germanhood, even if second-class, of identification with one's state no matter what, was probably what motivated Cohn in September 1933, when he wrote: 'I love Germany so much that the love is not diminished even by all the harassment we experience. Germany is the country whose language we use and where we have also had good days! One has to be loyal enough to accept even a government originating from a completely different camp.' In November 1941, German citizen Dr Willy (in honour of Kaiser Wilhelm) Cohn, holder of the Iron Cross from World War I, having surrendered his property to the State, was deported with his family, wife and two little daughters, and an entire transport of others to Kaunas in Lithuania. Once there, they formed even lines outside the walls of Fort IX, along pits resembling infantry trenches. And there they remained.

**Under the volcano**

The Germany of the early twentieth century resembled a volcano shortly before an eruption. The accumulated intellectual potential, industrial achievements, organizational efficiency, and growing wealth spawned a feeling of strength inevitably poised to be transformed into a real, tangible success. What stood in the way was the traditional balance of influences inherited from the nineteenth century and founded on the dominance of the major colonial powers. A sense of unsatisfied longing and a craving for change at all costs quickly gripped not only the masses but also a large part of the intellectual elite. Level-headed and widely respected at the age of 58, Max Planck, a future Nobel Prize winner, wrote in a November 1914 letter to Wilhelm Wien: 'Besides much that is horrible, there is also much that is unexpectedly great and beautiful: the smooth solution of the most difficult domestic political questions by the unification of all parties ..., the extolling of everything good and noble.'[18] Earlier, in the first weeks of the war, in September 1914, Planck enthusiastically wrote to his sister: 'What a glorious time we are living in. It is a great feeling to be able to call oneself a German.'[19] Awareness of the necessity for the nation to undergo a short but intensive purification process, similar to the tempering of a steel cast, was shared almost universally.[20] Sceptical about the idea of a 'holy fire', from which a new German nation was to emerge, were not only the social democratic circles but also by some financiers and industrialists, who could not see the point of rashly entering risky and uncertain war projects. The eruption of the volcano could not be stopped.

---

[15] 'Der Eintritt in eine große Nation kostet seinen Preis; die Hannoveraner und die Hessen und wir Schleswig-Holsteiner sind daran ihn zu bezahlen, und wir fühlen es wohl, daß wir damit von unserem Eigensten ein Stück hingeben. Aber wir geben es dem gemeinsamen Vaterland. Auch die Juden führt kein Moses wieder in das gelobte Land.' see note 13

[16] Fritz Haber – lecture at the invitation of the Medical Faculty of Berlin University, 16 June 1926, published in Fritz Haber, *Aus Leben und Beruf: Ansätze, Reden, Vorträge* (Berlin, Julius Springer, 1927)

[17] Willy Cohn, Norbert Conrads (Ed.), *Kein Recht, nirgends. Tagebuch vom Untergang des Breslauer Judentums 1933-1941*, (Köln, Böhlau Verlag, 2007)

The reviving tempering bath, *Stahlbad,* quickly turned into a *Blutbad.* It turned out to consist of hopeless burrowing in the ground churned up by artillery fire, bristled with barbed wire, and enveloped in the sickening odour of decaying bodies. What had been heralded as a *Blitzkrieg,* a lightning war, turned out to be a trench war of attrition. Germany found itself on the brink of disaster. The resources and the production capacity of ammunition factories, sufficient for a short lightning war planned for by the General Staff, proved completely inadequate for the purposes of the ongoing conflict. The British blockade successfully stopped transports of Chilean saltpetre used for the production of explosives. As early as September 1914, a team of experts was appointed to find a way out of the technological trap that Germany, fighting on two frotrnts, had found itself in. The team included Fritz Haber, who was already a *Geheimrat,* or privy councillor, director of the Kaiser Wilhelm Institute, an establishment whose primary goal was to catch up with, and get ahead of, American research institutes. He rubbed shoulders with members of the Berlin government circles and was a brilliant scientist and a splendid organizer with a natural ability to put together efficient implementation teams.

Fritz Haber considered it his civic duty to contribute to the war effort of his fatherland, the more so that because of his specialization and position he was part of the core industrial/scientific circle. The technological process he had developed, originally for the synthesis of ammonia, made it possible to close the ammunition gap after the requisite modifications and upgrades to the BASF chemical works at Leuna and Oppau. That, however, would not be enough to gain a distinct advantage and win the war. It was impossible to win quickly by only increasing the firepower of rifles and cannons and sending additional divisions to the front. Haber concluded that a quick victory was only feasible if the war was given a new technological dimension, introducing an element of shock and terror, moving outside the traditional patterns of general staff thinking. Such a shock was to be caused by the use of chemical weapons on the battlefield. What he had in mind was not any of the various irritant gases that force enemy troops to get out of the trenches straight under machine gun fire. Haber expected a shock reaction to be brought about by true chemical weapons, not just

---

[18] J. L. Heilbron, *The Dilemmas of an Upright Man: Max Planck and the Fortunes of German Science* (Cambridge, Mass., Harvard University Press, 2000)
[19] *Ibid.*
[20] Joachim Radkau, *Das Zeitalter der Nervosität: Deutschland zwischen Bismarck und Hitler* (München, Hanser, 1998)

incapacitating but lethal gases, all-pervading, unstoppable, and leaving a free way for advancing German troops. As a chemist, he realized with full clarity that chemical weapons would never remain the exclusive domain of one side. For that reason, he attached so much weight to the element of surprise, shock, and a spectacular military success, on a scale that would force the enemy to capitulate. He argued that a quick victory achieved in that way would on the whole reduce losses and save human victims on all sides. The Hague conventions in effect at that time prohibited the use of projectiles filled with asphyxiating gases, but Haber quickly convinced himself and others that if something was technologically possible it would definitely be used if only it could make more likely to win the war. The threat of possible secret French and English research on chemical weapons turned out to be an effective and decisive argument against those opposing the violation of existing conventions.

As a matter of fact, that line of reasoning did not differ much from the American Manhattan project during World War II. The use of nuclear weapons in Japan certainly caused a shock, helped to bring forward the end the war, already won anyway, and saved many lives, at least American ones. In contrast to Haber's old-fashioned ideas, the new weapon was used not on the battlefield but against civilian population, in agreement with the generally accepted military doctrine of World War II. The amazing effectiveness of the weapon intensified the shock experienced by Japan's staff officers, prompting capitulation, which was completely contrary to the Japanese war tradition.

Fritz Haber turned the Kaiser Wilhelm Institute he headed into an efficient machine supporting the ongoing war effort. The task of the institute for the time of war was not only to develop new, more effective kinds of chemical weapons but also to devise adequate protections, in reasonable anticipation of imminent use of similar weapons by the opponents. The institute grew immensely: 1,500 people, including 150 scientists, worked, or rather served, there. They included future Nobel Prize winners, such as Otto Hahn, James Franck, and Gustav Hertz. The scale of engagement and the organizational conception developed by Haber foreshadowed a new era of direct involvement of science in war, portending the Manhattan project a war and a generation later. Similar establishments were also brought into existence on the Entente side, especially in England and France.

Even though it is estimated that nearly half of the shells used in the last year of the war were filled with war gases, chemical weapons were in no way decisive with regard to the final outcome of World War I. Used by all sides, they were the cause of individual tragedies of soldiers but had no effect on the plans of army staffs. A representative of the United States, refusing to sign a Hague declaration banning chemical weapons, effectively stated that there was little difference between the allowable use of a stream of molten metal tearing apart human bodies and the illegitimate use of poisonous gas filling the lungs.[21] Haber's chlorine released directly from thousands of cylinders towards the trenches at Ypres was only a prelude. It was the subsequent generations of chemical weapons, developed after the war, and the improved means of delivery that led to the creation of a certain balance of fear, effectively blocking the use of chemical warfare during World War II.

---

[21] *Peace Conference at the Hague 1899: Report of Captain Mahan to the United States Commission to the International Conference at the Hague, on Disarmament, etc., with Reference to Navies,* http://avalon.law.yale.edu/19th_century/hag99-06.asp

**Death on the frontline, death in the extermination camps**

Judging past events by the present standards can be tricky. The traumatic experiences of World War II shaped the historical awareness of post-war European generations for decades; they carved mental furrows that continue to channel diverse streams of thoughts, sometimes without their true origins being realized. The figure of Fritz Haber, viewed through the conceptual filters of the Holocaust, mass extermination, fallen totalitarian systems, and additionally framed in contemporary political correctness, becomes a bizarre construct that has little to do with the realities of his day. Such chains of associations as Haber–Zyklon B–Auschwitz–Holocaust or Haber–Poison Gases–Mass Extermination–Dr Death, used in cheap journalism, inevitably lead into the wilderness of populism. They distort Haber into a forerunner of Dr Mengele from the Auschwitz ramp, or, at best, a grotesque mass-murder maniac à la Dr Strangelove from Stanley Kubrick's cult film.[22] Following this path, Haber could also be held responsible for acid rain, deforestation, and desertification of Africa, based on the chain of associations: Haber–Ammonia–Artificial Fertilizers–Environmental Pollution. Zyklon B, existing in the public mind as a tool of mass extermination, was developed as a strong and effective insecticide. As a matter of fact, it is still used as such but, to avoid grim associations, marketed under a different name. It is produced as Uragan D2 at a plant in Kolin, Czech Republic.[23] In fact, it is the same plant that, then known as Kaliwerken, supplied Zyklon B for Auschwitz in 1943–1945.

**Taboo**

Even without the modern, ahistorical associations linking the battlefields of the Great War with the gas chambers of Auschwitz, the use of war gases and similar chemicals has been taboo for a long time. Lethal chemicals were inescapably associated with poisons, regarded as treacherous, despicable, and cruel in Western culture. It is no coincidence that the first treaty on chemical warfare was the Strasbourg Agreement of 1675 signed between France and the Holy Roman Empire of the German Nation.[24] Both parties agreed to refrain from using poisoned bullets against each other. The agreement was entered into between the flagging German Empire of Leopold I and Louis XIV's France at the height of its power. Two hundred and forty years later a hardened German Reich of Wilhelm II released a chlorine cloud against the French at Ypres.

   The European taboo against chemical warfare was reflected in several international treaties signed during the industrial ear in Europe. The Hague Convention with respect to the Laws and Customs of War on Land signed in 1899 prohibited the signatory states from using poisoned arms in hostilities between them, and the accompanying Hague Declaration of 29 July 1899 banned the use of projectiles the object of which was the diffusion of asphyxiating or deleterious gases. The World War I belligerents tried to evade the prohibition of the declaration from the very start of the armed conflict.[25] Gases were to serve primarily as a means to help overcome the stalemate of trench warfare. France used grenades filled with tear gas as early as August 1914, to which Germany responded with a heavy mortar shell

---

[22] Stanley Kubrick (dir.), *Dr. Strangelove or How I Learned to Stop Worrying and Love the Bomb*, film, 1964
[23] Lučebni závody Draslovka a.s. Kolin
[24] Kim Coleman, *A History of Chemical Warfare* (Basingstoke, Palgrave Macmillan, 2005)
[25] Jonathan Tucker, *War of Nerves: Chemical Warfare from World War I to Al-Qaeda* (New York, First Anchor Book, 2007)

combining shrapnel bullets with a gas irritant. Both of these attempts, and some similar ones, passed almost completely unnoticed by those attacked as the effectiveness of the chemical agents in those *ad hoc* projectiles was close to zero.

Fritz Haber and his team approached the issue using full scientific methodology, carrying out detailed analyses, plotting mortality curves for various laboratory animals, chiefly cats, mice, and dogs, and analysing different atmospheric conditions. The first chlorine gas attack at Ypres was a complete success from a technological point of view, whereas militarily it was a local episode, and ethically it was a violation of the taboo against poison death in the name of a short-term tactical benefit. The Allies, having appropriately expressed their indignation, embarked on the organization of their own chemical forces, now with full scientific support. Like the Germans, they established identically specialized laboratories and even chemical warfare testing grounds. In declassified documents published after the war, it was noted that one of the parameters measured at the German laboratories, the *Tödlichkeitsprodukt,* or lethal index, was markedly lower (indicating a more 'efficient' substance) than the corresponding index measured for the same substances at American laboratories.[26] With true scientific perspicacity, it was explained as a result of wartime malnutrition of German lab cats compared with their American counterparts.

As early as Autumn 1915, a new invention made its way to the battlefields: phosgene, the true superstar of the chemical warfare of that time. Known previously from the dye industry, it was used successfully for the first time as a chemical weapon by France thanks to the inventiveness of Victor Grignard, the French winner of the 1912 Nobel Prize in chemistry. Ten times more effective then chlorine, it accounted for about 80 per cent of the deaths caused by chemical weapons during World War I.

**Clara**

Haber's black legend would not be so black were it not for Clara Immerwahr and her suicide. As with Zyklon B, today's criteria informed by knowledge from a later time try to make her primarily a victim of Haber's almost military despotism that not only stifled her career but betrayed the mission of science itself. The marriage had its epilogue on 1 May 1915 at Dahlem, on the night of a party that Fritz Haber, director of the Kaiser Wilhelm Institute, gave to celebrate his last great success: promotion to the rank of captain. It was one more dream come true, all the more precious as the promotion was granted outside the usual procedure in recognition of his outstanding merit. It was also a triumph which proved to be the last drop that caused Clara's cup of bitterness and disappointment to overflow.

While today's perspective of decades of peace and equal rights for women seems obvious, a picture of Clara composed only of elements presenting her as an innocent victim of her husband appears to be a product of moral reductionism, neglecting the context and the richness of nuances in history. In contrast to the conditions in which Marie Skłodowska-Curie was able to study and work, women in the German Empire did not publicly manifest dissatisfaction with their position. Neither the women's suffrage movement nor the opening of universities to women in Paris inspired similar initiatives in Germany. A climate or a model for the wife and mother pursuing a scientific career had yet to emerge. The Habers were probably the first university married couple, both holding doctorates *cum laude*. They had met

---

[26] Augustin M. Prentiss*, Chemicals in War: A Treatise on Chemical Warfare* (New York, McGraw-Hill, 1937)

as students at university, members of the peculiar research community that had only just begun to establish its social and professional codes and was completely unaccustomed to the presence of women. Clara agreed to marry Fritz after no fewer than ten years, when he already had an established scientific position at Karlsruhe, and she had begun her career at professor Richard Abegg's university laboratory in Breslau. Following an initial period of collaboration, in particular after the birth of their son, Fritz succumbed to the traditional model of family and stopped caring for Clara's scientific needs. Fritz's and Clara's characters and ambitions did not turn out to be complementary, which led to frequent clashes, as Clara confessed in a letter to Abegg:[27]

> What Fritz has gained during these eight years, I have lost ... Even if external circumstances and my own peculiar temperament are to blame too, the largest share of responsibility for this loss rests with Fritz and his permanent self-confidence and certainty about his place in the marriage and in running the household. He simply destroys every personality that is unable to stand up to him, like me. I keep asking myself if superior intelligence is sufficient to be a more valuable human being than others and if the part of me that has gone to the devil only because I did not meet the right man was not more important than even the most significant part of the theory of electrons.

A few years later, when the war broke out, the differences between them grew even deeper as a result of conflicting views on the question of using chemical weapons; the discord reached its climax and ended in a tragic gunshot. It was not the only suicide in the family, however. Aside from Clara's cousins, one cannot omit to mention the suicide of the Habers' son Hermann and then one of his daughters. Fritz Haber himself, accused of callousness because he left the same morning for the front, made the following confession in a letter to Karl Engler, his former rector at Karlsruhe, six weeks after Clara's death:

> I did not know if I would survive this month. But the war, full of its horrible sights and constantly requiring all my strength, has been able to soothe me. I was lucky to spend eight days working at the ministry, so I had a chance to see my son. Now I'm back at the frontline. Working amid the wartime complications, amongst unfamiliar people, I have absolutely no time to rest, reflect, or delve into my own feelings. The only thing left is concern about my stamina: will I be able carry the burden that has been put on my shoulders? … Every next day of bullets whizzing past is good for me. Here, only the present moment counts ... But when I get back to the staff office, clinging to the telephone receiver I can still hear in my heart the words she once told me and, exhausted, I can see her head appearing amid the orders and cablegrams, and it gives me pain ...'[28]

The tragic end of this marriage also marks another transition: a transition to the very heart of modern times. On the one hand, Fritz Haber's patriotism, his steadfast desire to win the war, led him to violate the old taboo mentioned above in the name of effectiveness of action. That became a manifestation of the solidifying industrial era, when faith in the power of intelligence and admiration for the power of manufacture brought about moral relativism. Efficiency and speed of action imperceptibly became a more important value than ethics. On the other hand, new, different needs were emerging on the part of women with academic ambitions. In the German society of the early twentieth century, Clara, a doctor of chemistry, found herself walking a pioneering, lonely road to what would be a different world, a world that would accept and similarly appreciate competence but which would require partnership and closeness, a different emotional quality of collaboration. Meanwhile, the relationship over time became increasingly a confrontation of two models: Haber belonged to the vanishing world of male authority that imposed order on the world and gave it meaning. To Clara that world became completely senseless that night, remaining only a victorious absurdity of men celebrating their efficiency and plunging into war.

---

[27] Clara to Abegg, 25 April 1909, *Archives of the Max Planck Society,* Haber Collection Rep. 13, 812
[28] Haber to Engler, fieldpost, June 1915, *Archives of the Max Planck Society,* Haber Collection, Rep. 13, 985

**Fritz Haber – one of us?**

The conviction of the victorious Allies that Haber's direct involvement in the production and application of chemical weapons should disqualify the inventor of ammonia synthesis from being considered for the Nobel Prize did not impact on the evaluation of his pre-war achievements. The issue of awarding the Nobel Prize to a scientist who would have achieved a breakthrough in the production of nitrogen-based artificial fertilizers was on the agenda of the Nobel Prize committee since 1909; only a suitable candidate was needed.[29] When the committee announced its first post-war nominations, it fully recognized the importance of Haber's work and, following years of debating, honoured the man who had discovered a method for using atmospheric nitrogen. The awards ceremony proceeded, not without ostentatious protests and conspicuous absences on the part of the Allies, as a kind of reminder of the causes of the great conflict: in addition to Haber, physics prizes for 1918 and 1919 were also awarded to German scientists (Max Planck and Johannes Stark). The decision on the award of the Nobel Prize in chemistry for 1918 was made when World War I was still a recent memory rather than just one more event in world history. Haber's role and personal involvement in the work on poison gases were no secret to anyone, and the victorious Allies indicated they would compile a list of war criminals – which was actually never published – that might also include the originator and patron of German chemical weapons. Nevertheless, the Nobel Prize committee, fulfilling the directive of Alfred Nobel's testament, recognized Haber not as the creator of chemical weapons but as one of those who had 'conferred the greatest benefit on mankind'. The just-ended war was mentioned only once in the presentation speech by Åke G. Ekstrand, the then President of the Royal Swedish Academy of Sciences: '[T]he protracted World War has sufficiently demonstrated to every country the need of organizing, wherever possible, production of essential commodities within its own borders in sufficient quantities to meet its own needs.'[30]

Fritz Haber, a grand master of technocracy, perceived the world as a series of technological problems to be solved. In the same spirit in which he solved the problem of binding atmospheric nitrogen and desired to tip the scales during World War I, he then attempted to help his fatherland faced with the requirement to pay murderous reparations imposed by the Treaty of Versailles. Haber's project to extract gold from sea water, pursued from 1920 to 1926, and involving several ocean voyages, ended in failure, as the concentration of gold was found to be much too low: the cost of extracting the trace quantities of gold found would exceed the gold value. It is also possible that Haber took to the oceans driven by the genetic heritage of his mother's adventurous brothers, Ludwig and Edward Haber of Brieg (today Brzeg) in Lower Silesia, the sons of a quiet and pious Jewish merchant importing wool from Poland and grain from Russia. Edward is known to have been a merchant and a consul in San Salvador. Ludwig travelled the length and breadth of the world: West Africa, Egypt, India, Ceylon, Java, and China, to finally reach Japan. There, at Hakodate in 1874, the 32-year-old Ludwig was hacked to pieces by a samurai consumed with hate for white foreigners.[31]

---

[29] Elisabeth T. Crawford, *The Beginnings of the Nobel Institution: The Science Prizes, 1901-1915* (Cambridge University Press, 1987)
[30] http://nobelprize.org/nobel_prizes/chemistry/laureates/1918/press.html

Fritz Haber was unable to give up his technocratic attitude even in his personal life. He felt best in his professional setting – a good organizer, always full of ideas, protective towards his co-workers. However, what worked well in the laboratory conditions of research institutes was not necessarily best for the realities of family life. Two unhappy marriages, one ended by the suicide of his wife, the other, with Charlotte, taking place in two incompatible worlds, as it were, and leading to a split after ten stormy years.

Fritz Haber, a technocrat who endeavoured throughout his life to cultivate the traditional *Junker* virtues, a Prussian nationalist who was nevertheless free of contempt for others – who was he? How should he be judged? His tumultuous life was far from being a life of academic routine; it was no less eventful than the adventure-novel life of his uncle murdered in Japan. The latter died at the hands of a Japanese nationalist hateful of foreigners; the former was crushed by the totalitarian Nazi system, a lunacy of the nation that boasts the heritage of Goethe, Beethoven, and Kant. In a manner typical of technocrats, Fritz Haber enthusiastically believed in simple solutions to complex problems. Ammonia synthesis, which brought him fame, was a matter of setting appropriate reaction conditions and finding the right catalyst. Winning the war was to be a matter of using the right chemicals on the battlefield. Similarly, gold from sea water was to help Germany meet the contributions imposed by the Treaty of Versailles.

Nazi state-licensed anti-Semitism was not amenable to interpretation compatible with technological rationality. In one of his last letters, Haber, already in exile in England, wrote the following to Bosch: 'I never did anything, never said even a single word, that could warrant making me an enemy of those now ruling Germany.'[32] The millions of Germans who did not utter a single word of objection or make the slightest gesture of protest contributed to the rise of Hitler's state. The philosophy of the state that Haber had enthusiastically professed just a few years before, boiling down to the technological recipe: *'*Our state does not serve its citizens, but the citizens the state,' revealed its limitations.

Considered from today's perspective, the figure of Haber is susceptible to easy manipulation. It is much easier to reduce his achievements to war gases; that is something almost everybody will understand. It is harder to deal with the accomplishments that earned him a Nobel Prize. We live convinced of our own uniqueness, uniqueness as individuals and the uniqueness of our times. Fed slogans about all-pervasive globalization, information revolution, and an extraordinary rate of change, we look indulgently at the lazy flow of time in the centuries past. In reality, however, the seemingly smooth flow is only an illusion arising from the distance as much as from a lazy reluctance to get closer. The world of Fritz Haber changed dramatically over the 66 years of his life, between 1868 and 1934, certainly no less than our modern world has changed over a similar period, since the end of World War II.

Emotional disputes over the assessment of contemporary figures are summed up using the convenient phrase 'history will judge', as if history were a kind of developer bringing out a latent picture on exposed photographic film. It is impossible to completely take off the glasses of the present when looking at the past. If it were possible, history would become a dead discipline of learning, fixed in a form given to it once and for all. Any description must also be an understanding. One has to understand the motives, the aims, and the consequences of actions and to comprehend the existing conditions, circumstances, and constraints. An account emerging in our contemporary times becomes like a musical piece composed a long time ago: played on a modern instrument, exclusively on the basis of flat musical notation, it becomes defective and incomplete without familiarity with the period, the composer, his achievements and his intentions.

---

[31] Peter Fraenkel, *Ludwig Haber – The Consul and the Samurai*; http://ludwighaber.blogspot.com/
[32] Haber to Bosch 28 December 1933, *Archives of the Max Planck Society,* Haber Collection, Rep. 13, 911

Fritz Haber – our contemporary, ever more distant in the flux of time, still stirring emotions.

**Acknowledgments:** We are grateful for comments and suggestions kindly provided to us by Gerhard M. Oremek (Frankfurt), Bretislav Friedrich (Berlin), Adam Jezierski (Wroclaw)

**Post scriptum**

This essay is being written as a French film crew is shooting a documentary in Wrocław, retracing the footsteps of Fritz Haber and his family, with the participation of Fritz and Clara's granddaughter, Isabelle Traeger. The documentary is not so much about Haber himself but primarily about the passion of David Vandermeulen of Brussels, captivated by Haber's figure.

Wrocław, 20 August 2011